\documentclass[english,12pt,a4paper]{article}

\usepackage{authblk}
\usepackage[text={17.0cm,23.7cm}]{geometry}
\usepackage[T1]{fontenc}
\usepackage[utf8]{inputenc}
\usepackage[english]{babel}
\usepackage{graphicx}
\usepackage{amsmath,amssymb}
\usepackage{siunitx}
\usepackage[dvipsnames]{xcolor}
\usepackage{yfonts}
\usepackage[normalem]{ulem}
\usepackage{enumitem}
\usepackage[mathscr]{euscript}
\usepackage{float}
\usepackage{csquotes}
\usepackage[numbers,sort&compress]{natbib}
\usepackage{booktabs}
\usepackage{hyperref}
\usepackage{placeins}
\newcommand{\Msun}{M_\odot}
\newcommand{\Rsun}{R_\odot}
\newcommand{\Lsun}{L_\odot}
\newcommand{\mesa}{\textsc{mesa}}
\newcommand{\MD}{M_{\rm D}}
\newcommand{\Mb}{M_{\rm b}}
\newcommand{\RD}{R_{\rm D}}
\newcommand{\Teff}{T_{\rm eff}}
\newcommand{\Tc}{T_{\rm c}}
\newcommand{\nablarad}{\nabla_{\rm rad}}
\newcommand{\nablaad}{\nabla_{\rm ad}}

\begin{document}

\title{\vspace{-1.4cm}\textbf{Stellar structure with compact dark matter cores}\\[8mm]}
\author{Alejandro Ibarra}
\author{Halim Shaikh}
\affil{\normalsize\textit{Technical University of Munich, TUM School of Natural Sciences, Physics Department,\\ 85748 Garching, Germany}}
\maketitle

\begin{abstract}
We investigate the structure of stars subject to the gravitational effects of compact dark matter cores at their centers. We first study the problem analytically using the homology approximation and then perform a more detailed numerical analysis using the stellar evolution code \mesa{}. We find that a compact dark core can significantly change the properties of a star and its position in the Hertzsprung-Russell diagram. In particular, the presence of a sufficiently massive dark core can cause the star to expand significantly, becoming more luminous and having a lower surface temperature. As a result, a star with the chemical composition of a main-sequence star can acquire the appearance of a red giant.
\end{abstract}

\section{Introduction}\label{sec:intro}

The structure of a main-sequence star is determined by the balance between gravity, pressure, nuclear energy generation, and energy transport (see, {\it e.g.},~\cite{Kippenhahn2012}). Its position in the Hertzsprung-\allowbreak{}Russell (HR)~\cite{Hertzsprung1911,Russell1914} diagram depends mainly on its mass and chemical composition. However, standard models of stellar structure generally neglect the possible presence of dark matter, which constitutes a substantial fraction of the matter content in our Universe and, in particular, in our Galaxy \cite{Planck:2018vyg,Bland_Hawthorn_2016,Cautun:2019eaf}.

If dark matter consists of particles that scatter sufficiently efficiently with protons or electrons, they can be captured by stars and eventually accumulate in their central regions, where they may thermalize with the stellar plasma~\cite{Press:1985ug,Gould:1987ir,Gould:1987ju}. These particles can transport energy from the stellar interior to the outer layers, potentially modifying the stellar structure and its position in the HR diagram~\cite{Spergel:1984re,Gould:1989hm}. Furthermore, if dark matter particles annihilate into Standard Model particles, the additional energy released may also affect the stellar equilibrium~\cite{Spolyar:2007qv,Freese:2008wh,Scott:2008ns,Iocco:2008rb}.

In this paper, we focus on a different possibility, namely that dark matter is asymmetric and strongly self-interacting, in analogy with protons and nuclei in the visible sector. If this is the case, stable macroscopic configurations of dark matter could form~\cite{Colpi:1986ye,Kouvaris:2015rea}. In the early Universe, dark matter and visible matter could coexist within the same overdense region and form a compact dark object while the visible matter remains relatively diffuse. The gravitational potential generated by this object could then act as a seed for the subsequent formation of a visible star, eventually leading to a stellar configuration with a compact dark core at its center. The additional gravitational field produced by the dark core would modify the hydrostatic equilibrium of the surrounding stellar material, potentially shifting the star away from its conventional position in the HR diagram.

The work is organized as follows: we investigate in Section~\ref{sec:homology} the effects of the additional gravitational potential using the homology approximation, which allows us to derive analytical scaling relations for the stellar properties. In Section~\ref{sec:mesa}, we then perform a numerical analysis using the stellar evolution code \mesa{}, examining how the presence of the dark core modifies the stellar radius, luminosity, and effective temperature, and hence its position in the HR diagram. Finally, in Section~\ref{sec:conclusions}, we present our conclusions.

\section{Stellar structure with a dark core: homology approximation}
\label{sec:homology}

We consider a star of baryonic mass $\Mb$ with a compact dark core of mass $\MD$ and radius $\RD$. The dark matter interacts with the stellar gas only through gravity, without participating in nuclear reactions or exchanging heat with ordinary matter. Nevertheless, its gravitational field modifies the stellar pressure and temperature profiles, and hence nuclear energy generation and transport. We first study these effects using the homology approximation.

For a stationary, spherically symmetric configuration, the equations governing the baryonic matter are \citep{Kippenhahn2012}
\begin{align}
 \frac{dm_b}{dr}&=4\pi r^2\rho,\nonumber\\
 \frac{dP}{dr}&=-\frac{G\rho}{r^2}\bigl[m_b(r)+M_D(r)\bigr],\nonumber\\
 \frac{dL}{dr}&=4\pi r^2\rho\,\epsilon_{\rm nuc},\nonumber\\
 \left.\frac{dT}{dr}\right|_{\rm rad}
 &=-\frac{3\kappa\rho L}{16\pi a_{\rm rad}c\,r^2T^3}.
 \label{eq:structure}
\end{align}
Here $m_b(r)$ and $M_D(r)$ denote the baryonic and dark masses enclosed within radius $r$, respectively, while $\rho$, $P$, $T$, and $L$ are the density, pressure, temperature, and luminosity of the stellar gas. Furthermore, $G$ is Newton's gravitational constant, $\kappa$ the opacity, $\epsilon_{\rm nuc}$ the nuclear energy-generation rate per unit baryonic mass, $a_{\rm rad}$ the radiation constant, and $c$ the speed of light. The last equation applies to radiative energy transport; in convective regions, the temperature gradient must be determined separately. The dark core enters these equations explicitly only through gravity, but its gravitational field also modifies the density and temperature profiles, thereby affecting nuclear burning and energy transport.

Since the enclosed baryonic mass is small near the center, the dark core may dominate the gravitational field in the nuclear-burning region even when $\MD\ll\Mb$. It may also change the boundaries between radiative and convective regions. Combining the hydrostatic and radiative temperature gradients in Eq.~\eqref{eq:structure}, we obtain
\begin{equation}\label{eq:schwarz}
 \nablarad\equiv\left(\frac{d\ln T}{d\ln P}\right)_{\rm rad}
 =\frac{3\kappa LP}{16\pi a_{\rm rad}cG[m_b(r)+M_D(r)]T^4}.
\end{equation}
According to the Schwarzschild criterion, a region is convectively unstable when $\nablarad>\nablaad$, with $\nablaad\equiv(d\ln T/d\ln P)_S$. The dark core may therefore affect not only the location of nuclear burning but also the extent of the convective regions.

To derive approximate scaling relations, we assume an ideal-gas equation of state and power-law expressions for the opacity and nuclear energy-generation rate,
\begin{equation}\label{eq:powlaws}
 P\propto\rho T,\qquad
 \kappa=\kappa_0\rho^aT^b,\qquad
 \epsilon_{\rm nuc}=\epsilon_0\rho^sT^\nu.
\end{equation}
We define the dark-to-baryonic mass ratio $f\equiv\MD/\Mb$ and the total gravitational mass $M_g=\Mb+\MD=\Mb(1+f)$. In the homology approximation, the characteristic density and temperature scale as
\begin{equation}\label{eq:scales}
 \rho_{\rm h}\propto\frac{\Mb}{R^3},\qquad
 T_{\rm h}\propto\frac{GM_g}{R},
\end{equation}
where $R$ is the stellar radius. These relations describe the global response but do not capture the stronger gravitational field produced by a compact dark core near the center.

To describe this local effect, we approximate the dark core by a uniform-density sphere with $\RD\ll R$. Its gravitational potential, taken to vanish at infinity, is
\begin{equation}\label{eq:phi}
 \Phi_D(r)=-\frac32\frac{G\MD}{\RD}\,
 \varphi\!\left(\frac{r}{\RD}\right),\qquad
 \varphi(x)=\begin{cases}
 1-x^2/3,&x<1,\\
 2/(3x),&x\geq1.
 \end{cases}
\end{equation}
We assume that the central region is approximately adiabatic, as expected for the convective core of the reference star. Hydrostatic equilibrium, the ideal-gas equation of state, and the definition of $\nablaad$ then give
\begin{equation}
 \frac{dT}{dr}=-\nablaad\frac{\mu m_u}{k_{\rm B}}\frac{d\Phi}{dr}.
\end{equation}
Here $\Phi=\Phi_b+\Phi_D$ is the total gravitational potential, including the contributions from baryonic and dark matter, $\mu$ is the mean molecular weight, and $m_u$ is the atomic mass unit. Integrating the contribution from the dark potential and neglecting its value at the outer matching radius gives
\begin{equation}
 \Delta T(r)\simeq-\nablaad\frac{\mu m_u}{k_{\rm B}}\Phi_D(r).
\end{equation}
Using Eq.~\eqref{eq:phi}, the approximate temperature profile becomes
\begin{equation}\label{eq:localT}
 T(r)\simeq T_{\rm h}\left[1+\lambda\varphi(r/\RD)\right],
\end{equation}
where
\begin{equation}\label{eq:lambda}
 \lambda=\frac32\nablaad\frac{\mu m_u}{k_{\rm B}}\frac{G\MD}{\RD T_{\rm h}}.
\end{equation}
Normalizing the homology temperature to the star without a dark core, $T_{\rm h}=T_{c,0}(1+f)R_0/R$, yields
\begin{equation}\label{eq:lambda0}
 \lambda=\lambda_0\frac{f}{1+f}\frac{R}{R_0},
 \end{equation}
 with 
 \begin{equation}
 \lambda_0=\frac32\frac{\nablaad}{c_T}\frac{R_0}{\RD}
 \end{equation}     
 and 
 \begin{equation}
 c_T\equiv\frac{k_{\rm B}T_{c,0}R_0}{G\Mb\mu m_u}.
\end{equation}
Here and below, the subscript $0$ denotes the reference star without a dark core. Equation~\eqref{eq:lambda0} shows that the temperature enhancement depends on both the dark-to-baryonic mass ratio and the compactness of the dark object. This expression relies on the assumption of an approximately adiabatic central region and becomes less reliable when the stellar structure departs significantly from that of the reference star.

For a monatomic ideal gas along an adiabat, $\rho\propto T^{3/2}$. We therefore approximate the local density profile by
\begin{equation}\label{eq:localrho}
 \rho(r)\simeq\rho_{\rm h}\left[1+\lambda\varphi(r/\RD)\right]^{3/2}.
\end{equation}

We now estimate the luminosity from nuclear burning and radiative diffusion. With $\epsilon_{\rm nuc}\propto\rho^s T^\nu$, the local enhancements in Eqs.~\eqref{eq:localT} and \eqref{eq:localrho} increase the nuclear power by a factor
\begin{equation}\label{eq:boost}
 B=\int dF_0\,\bigl[1+\lambda\varphi(r/\RD)\bigr]^{\nu+\frac32(1+s)},
\end{equation}
where $r=r_0 R/R_0$ with $r_0$ the radius of a shell in the reference star and
\begin{equation}\label{eq:dF0_definition_v2}
 dF_0=\frac{4\pi r_0^2\rho_0(r_0)\epsilon_{{\rm nuc},0}(r_0)\,dr_0}
 {L_{{\rm nuc},0}}.
\end{equation}
By construction, $\int dF_0=1$. Thus, $dF_0$ weights each shell by its contribution to the reference nuclear luminosity. Since this prescription does not determine the full response of the density and temperature profiles, $B$ is an approximate nuclear-burning enhancement. The luminosities inferred from nuclear burning and radiative diffusion scale as
\begin{equation}\label{eq:constraints}
 \begin{split}
 L_{\rm nuc}&\propto\Mb^{1+s}M_g^\nu R^{-3s-\nu}B,\\
 L_{\rm rad}&\propto\Mb^{-(a+1)}M_g^{4-b}R^{3a+b}.
 \end{split}
\end{equation}
Equating the two estimates at fixed baryonic mass gives
\begin{align}\label{eq:twozone}
 \left(\frac{R}{R_0}\right)^{\mathcal D}
 &=(1+f)^{\nu+b-4}B,\nonumber\\
 \frac{L}{L_0}&=(1+f)^{4-b}\left(\frac{R}{R_0}\right)^{3a+b},
\end{align}
where $\mathcal D\equiv3(a+s)+b+\nu$. The factor $B$ depends on both $R$ and $f$, so the first relation is generally solved numerically. The central temperature follows from  Eq.~\eqref{eq:localT}, while the effective temperature is obtained from the Stefan-Boltzmann relation $L=4\pi R^2\sigma_{\rm SB}T_{\rm eff}^4$, where $\sigma_{\rm SB}$ is the Stefan-Boltzmann constant:

\begin{align}
 \frac{\Tc}{T_{c,0}}&=(1+f)(1+\lambda)\frac{R_0}{R},
 \label{eq:twozoneT}\\
 \ln\frac{\Teff}{T_{{\rm eff},0}}
 &=\frac14\ln\frac{L}{L_0}-\frac12\ln\frac{R}{R_0}.
 \label{eq:TeffDeriv}
\end{align}
Consequently, the effective temperature of the star can decrease even if its central temperature and luminosity increase, provided that the stellar radius grows sufficiently.

For small dark core masses, expanding Eq.~\eqref{eq:boost} to first order in $f$ gives
\begin{equation}
 B\simeq1+\left[\nu+\frac32(1+s)\right]\eta\lambda_0f,
\end{equation}
where we have defined
\begin{equation}
 \eta\equiv\int dF_0\,\varphi(r_0/\RD).
\end{equation}
The corresponding first-order changes in the stellar properties can be expressed as
\begin{align}\label{eq:homscalings}
 \frac{R}{R_0}&\simeq(1+f)^{\alpha_R},\nonumber \\
 \frac{L}{L_0}&\simeq(1+f)^{\alpha_L},\nonumber \\
 \frac{\Teff}{T_{{\rm eff},0}}
 &\simeq(1+f)^{(\alpha_L-2\alpha_R)/4},
\end{align}
where
\begin{align}
 \alpha_R&=\frac{\nu+b-4+\left[\nu+\frac32(1+s)\right]\eta\lambda_0}
 {\mathcal D},\nonumber\\
 \alpha_L&=4-b+(3a+b)\alpha_R.
 \label{eq:exponents}
\end{align}
Therefore, the star expands for $\alpha_R>0$ and cools at its surface for $\alpha_R>\alpha_L/2$. 

For illustration, we consider a star with baryonic mass $\Mb=3\,\Msun$ and a dark core of radius $\RD=0.02\,\Rsun$, where $\Msun$ and $\Rsun$ denote the solar mass and radius, respectively. As a reference, we use the model without a dark core, computed with \mesa{} as described in Section~\ref{sec:mesa}. This model has radius $R_0=2.00\,\Rsun$, luminosity $L_0=79.8\,\Lsun$, effective temperature $T_{{\rm eff},0}=1.22\times10^4\,{\rm K}$, central temperature $T_{c,0}=2.40\times10^7\,{\rm K}$, and central density $\rho_{c,0}=39.5\,{\rm g\,cm^{-3}}$, where $\Lsun$ denotes the solar luminosity. The hydrogen mass fraction and mean molecular weight are $X=0.698$ and $\mu=0.618$, respectively. Taking $\nabla_{\rm ad}=2/5$, corresponding to a monatomic ideal gas, we obtain $c_T\simeq1.13$ and $\lambda_0\simeq53$, while the nuclear-burning profile gives $\eta\simeq0.1086$. The parameter $\lambda_0$ characterizes the temperature enhancement induced by the dark gravitational potential, whereas $\eta$ is the nuclear-luminosity-weighted average of its radial profile.

We adopt $s=1$, appropriate for an approximate hydrogen-burning rate per unit mass proportional to $\rho$ at fixed composition. We consider Kramers opacity, $(a,b)=(1,-3.5)$, and electron-scattering opacity, $(a,b)=(0,0)$, together with $\nu=4$ for pp burning and $\nu=16.7$ for CNO burning. Table~\ref{tab:homregimes} lists the resulting exponents. In all four cases, the small-$f$ approximation predicts that the star expands and cools. For finite dark masses, we solve Eq.~\eqref{eq:twozone} with the full factor $B$. The two treatments agree within about $1\%$ for $f<0.006$ in Cases B and D, and for $f<0.014$ in Cases A and C. This comparison measures the accuracy of the first-order expansion within the homology model, not its agreement with a self-consistent stellar calculation.

\begin{table}[t!]
\centering
\setlength{\tabcolsep}{5pt}
\begin{tabular}{@{}lccccccc@{}}
\toprule
Case & Opacity & Burning & $\alpha_R$ & $\alpha_L$
& $R/R_0$ & $L/L_0$ & $\Teff/T_{{\rm eff},0}$ \\
\midrule
A & Kramers & pp & $5.67$ &$4.66$ & $(1+f)^{5.67}$ & $(1+f)^{4.66}$ & $(1+f)^{-1.67}$ \\
B & Kramers & CNO & $6.40$ & $4.30$ & $(1+f)^{6.40}$ & $(1+f)^{4.30}$ & $(1+f)^{-2.12}$ \\
C & Electron scattering & pp &$5.77$ & $4.00$ & $(1+f)^{5.77}$ & $(1+f)^{4.00}$ & $(1+f)^{-1.88}$ \\
D & Electron scattering & CNO &$6.41$ &$4.00$ & $(1+f)^{6.41}$ & $(1+f)^{4.00}$ & $(1+f)^{-2.21}$ \\
\bottomrule
\end{tabular}
\caption{Homology exponents for different opacity and nuclear-burning laws in the limit of small $f$, assuming $\Mb=3\,\Msun$ and $\RD=0.02\,\Rsun$. The scaling relations are obtained to first order in $f$ and are therefore only valid for small dark core masses.}
\label{tab:homregimes}
\end{table}
As the star expands and its effective temperature decreases, it may approach the Hayashi limit. In this regime, the stellar envelope becomes strongly convective, and the approximation of a radiative envelope is no longer valid \citep{Kippenhahn2012}. To estimate the luminosity along the cool branch, we assume that most of the nuclear burning takes place within the dark core, where $m_b\ll\MD$. Assuming an approximately constant temperature gradient $\nabla=d\ln T/d\ln P$, we obtain from hydrostatic equilibrium in the gravitational potential of the dark core
\begin{equation}\label{eq:wellpressure}
 \frac{P_c}{\rho_c}\simeq\frac32\nabla\frac{G\MD}{\RD}.
\end{equation}
We take $\nabla=1/4$ and account for radiation pressure by introducing the fraction of the central pressure due to the gas, $\beta=P_{\rm gas}/P_c$. The central temperature is then given by $\Tc=\beta T_{\rm w}$, where
\begin{equation}\label{eq:giantstate}
 T_{\rm w}\equiv\frac38\frac{\mu m_u}{k_{\rm B}}\frac{G\MD}{\RD},
\end{equation}
and the central density is
\begin{equation}
 \rho_c=\frac{8a_{\rm rad}\Tc^4\RD}{9(1-\beta)G\MD}.
\end{equation}
Assuming that the opacity in the nuclear-burning region is dominated by electron scattering, $\kappa_{\rm es}=0.2(1+X)\,{\rm cm^2\,g^{-1}}$, we estimate the luminosity from radiative diffusion as
\begin{equation}\label{eq:giantL}
 L_{\rm rad}\simeq(1-\beta)\frac{4\pi Gc\MD}{\kappa_{\rm es}}.
\end{equation}
On the other hand, the luminosity from nuclear reactions within the dark core can be estimated as
\begin{equation}\label{eq:giantnuc}
 \begin{split}
 L_{\rm nuc}&\simeq\mathcal F\frac{4\pi}{3}\RD^3\rho_c\epsilon_{c,0}
 \left(\frac{\rho_c}{\rho_{c,0}}\right)^s
 \left(\frac{\Tc}{T_{c,0}}\right)^\nu,\\
 \mathcal F&=3\int_0^1x^2\left(1-\frac{x^2}{3}\right)^{\nu+3(1+s)}dx.
 \end{split}
\end{equation}
Here, $\rho_{c,0}$ and $\epsilon_{c,0}$ are the central density and nuclear energy-generation rate per unit mass of the star without a dark core, while the factor $\mathcal F$ takes into account the temperature and density profiles assumed in the nuclear-burning region. By imposing $L_{\rm rad}=L_{\rm nuc}$, we can determine $\beta$ and hence estimate the central temperature, central density and luminosity of the star. In our calculation, we take $\epsilon_{c,0}=1.5\times10^3\,{\rm erg\,g^{-1}\,s^{-1}}$ and $\mathcal F=0.059$ for CNO burning or $0.18$ for pp burning. Since our approximation does not describe the structure of the envelope, we assume a constant effective temperature of $T_{\rm H}=3700\,{\rm K}$ along the cool branch. In this regime, the luminosity depends mainly on the properties of the dark core, similar to what is found for ordinary red giants, where the luminosity is largely determined by the core mass \citep{Kippenhahn2012}.

\begin{figure}[t!]
 \centering
 \includegraphics[width=0.9\linewidth]{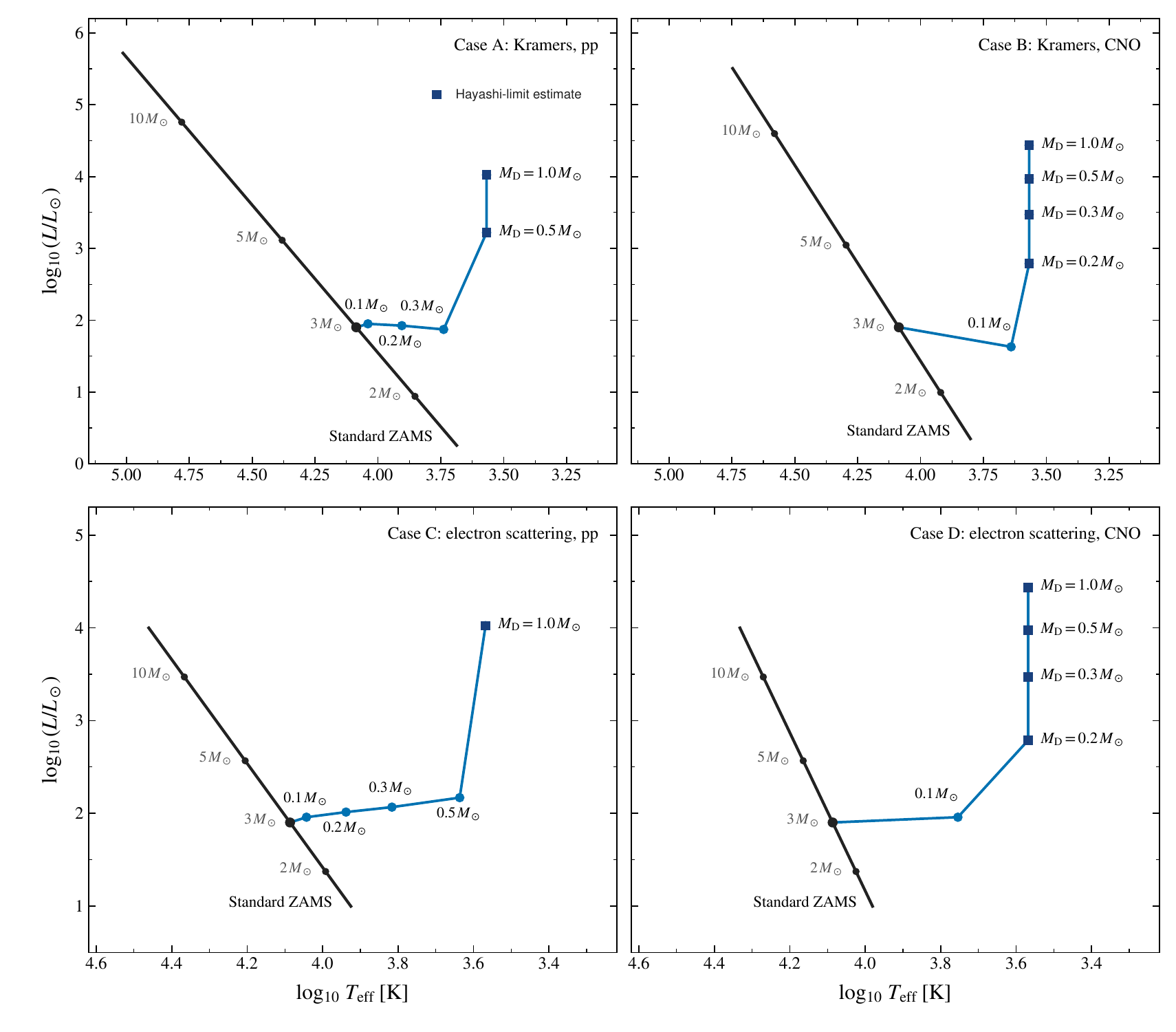}
 \caption{Illustrative displacement in the HR diagram for a star with $\Mb=3\,\Msun$ and $\RD=0.02\,\Rsun$, for the four opacity and burning prescriptions of Table~\ref{tab:homregimes}. The black curves show ordinary zero-age main sequences estimated by homology, and the markers indicate increasing dark core mass.}
 \label{fig:homhr}
\end{figure}

In Fig.~\ref{fig:homhr}, we show how the position of the star in the Hertzsprung--Russell diagram changes as the mass of the dark core increases. To construct the radiative branch, we use Eqs.~\eqref{eq:twozone} and \eqref{eq:TeffDeriv}, while for the cool branch we use Eqs.~\eqref{eq:giantstate}--\eqref{eq:giantnuc} and assume $\Teff=T_{\rm H}$. For comparison, we also show the zero-age main sequences obtained using the homology relations for stars without dark cores, $R_0\propto\Mb^{\beta_R}$ and $L_0\propto\Mb^{\beta_L}$, with $\beta_R=(a+b+s+\nu-2)/\mathcal D$ and $\beta_L=1+s+\nu-(3s+\nu)\beta_R$. In all cases, we normalize these relations to the properties of the reference star.

In all four cases, the star initially expands and its effective temperature decreases, while its luminosity changes by less than a factor of two. The radiative solution reaches $T_{\rm H}$ at $\MD\simeq0.11$ and $0.12\,\Msun$ in Cases B and D, and at $\MD\simeq0.43$ and $0.61\,\Msun$ in Cases A and C. On the cool branch, the luminosity rises with dark core mass, reaching about $2.7\times10^4\,\Lsun$ for CNO burning and $1.1\times10^4\,\Lsun$ for pp burning at $\MD=1\,\Msun$. The model therefore predicts that a compact dark core can produce an extended, cool, and luminous star while nuclear burning remains concentrated near its center.

The homology approximation illustrates how a compact dark core can modify the stellar radius, luminosity, and effective temperature. However, it relies on approximate interior profiles and separate treatments of the radiative and cool-giant branches. In the next section, we use numerical stellar-structure calculations to examine these effects without imposing the homology assumptions.

\section{Stellar structure with a dark core: numerical analysis}
\label{sec:mesa}

We now investigate in more detail how a compact dark core affects the structure of a star using the publicly available code \mesa{} \citep{Paxton2011,Paxton2013,Paxton2018,Paxton2019,Jermyn2023}. Unlike the homology approximation discussed in Section~\ref{sec:homology}, this numerical calculation allows us to determine how the physical quantities change inside the star and where energy is transported by radiation or convection. 

We consider a star with baryonic mass $\Mb=3\,\Msun$ and a spherical dark core with uniform density and fixed radius $\RD=0.02\,\Rsun$. As in the previous section, we assume that dark matter affects the star only through gravity. The dark matter mass enclosed within a radius $r$ is given by
\begin{equation}\label{eq:darkmassprof}
 M_D(r)=\MD\begin{cases}
 (r/\RD)^3,&r<\RD,\\
 1,&r\geq\RD.
 \end{cases}
\end{equation}
The baryonic gas extends all the way to the center of the star and is therefore also present inside the dark core. The radius $\RD$ simply determines the size of the region occupied by dark matter.

We increase $\MD$ from zero to $1\,\Msun$, corresponding to $0\leq f\equiv\MD/\Mb\leq1/3$, and calculate 3606 stellar configurations in hydrostatic and thermal equilibrium. We keep the chemical composition fixed, with a central hydrogen abundance of $X_c=0.698$. For all configurations with a nonzero dark mass, we require $|L_{\rm nuc}/L-1|<10^{-8}$. To obtain these configurations, we gradually increase $\MD$ and allow the stellar structure to adjust to the new value of $\MD$ at each step. The resulting configurations therefore describe a sequence of static equilibria rather than the evolution of a star over time. 

\begin{figure}[t!]
 \centering
 \includegraphics[width=\linewidth]{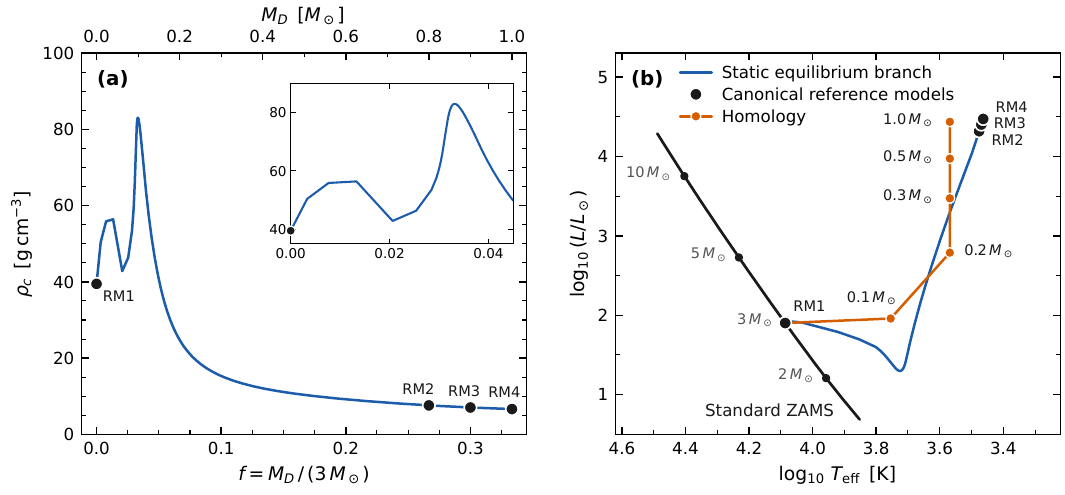}
 \caption{Properties of the equilibrium configurations obtained with \mesa{} for different values of the dark core mass. Panel (a) shows the central density, with an inset showing in more detail its dependence on $f$ at small values of $f$. Panel (b) shows the HR diagram. The black curve shows the ZAMS for ordinary stars calculated with \mesa{}, and the black circles indicate the four reference models RM1-RM4 (Table~\ref{tab:models}). The orange circles show the results of the homology approximation for Case~D of Table~\ref{tab:homregimes}. For the cool branch, we take $\Teff=3700\,{\rm K}$, as in Fig.~\ref{fig:homhr}.}
 \label{fig:mesahr}
\end{figure}

\begin{figure}[t!]
 \centering
 \includegraphics[width=\linewidth]{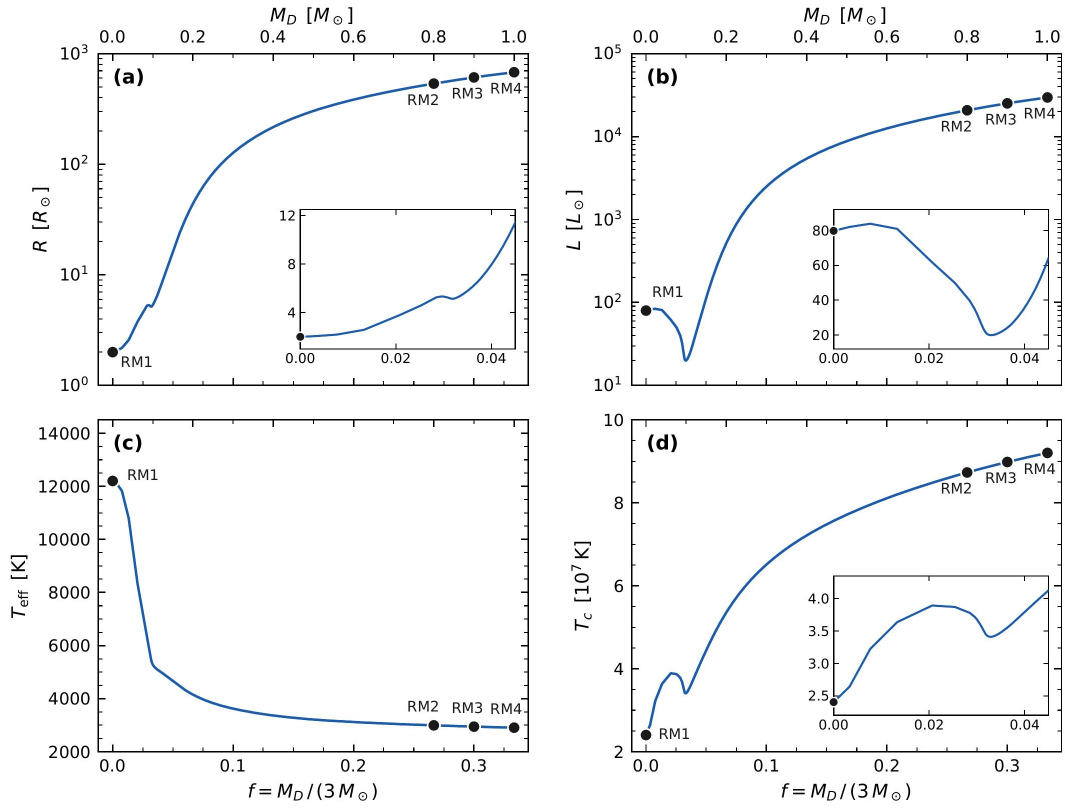}
\caption{Stellar radius, luminosity, effective temperature and central temperature obtained with \mesa{} as functions of $f=\MD/\Mb$. The black dots indicate the reference models RM1-RM4. The insets show in more detail how these quantities change at small values of $f$.}
 \label{fig:mesaglobal}
\end{figure}

Figures~\ref{fig:mesahr} and \ref{fig:mesaglobal} show the main results of our numerical calculation. For comparison, we also show in Fig.~\ref{fig:mesahr}(b) the zero-age main sequence (ZAMS) for stars without dark matter cores, with masses between $1.5$ and $15\,\Msun$. These configurations are calculated using the same physical assumptions in \mesa{} and the same central hydrogen abundance. We choose four reference models, RM1-RM4, corresponding to $\MD=0$, $0.8$, $0.9$, and $1.0\,\Msun$. These models are indicated in the figures, and their main properties are given in Table~\ref{tab:models}.

As $f$ increases, the star generally expands and its effective temperature decreases throughout the range considered. The luminosity, however, shows a more complicated behavior. It first reaches a small maximum at $f\simeq0.0077$, then decreases to about $20\,\Lsun$ at $f\simeq0.033$, and finally increases rapidly. The radius also shows a change in behavior around $f\simeq0.03$, after which the star expands considerably. For $\MD=1\,\Msun$, the stellar radius is about $679\,\Rsun$ and the luminosity is $2.96\times10^4\,\Lsun$, compared to $2.00\,\Rsun$ and $79.8\,\Lsun$ for the star without a dark core. The effective temperature decreases from $1.22\times10^4$ to $2.91\times10^3\,{\rm K}$, while the central temperature increases from $2.40\times10^7$ to $9.21\times10^7\,{\rm K}$. Therefore, a star with the chemical composition of a main-sequence star can have a radius, luminosity and effective temperature typical of cool giants or supergiants.

\begin{table}[t]
\centering
\begin{tabular}{@{}cccccc@{}}
\toprule
Model & $\MD/\Msun$ & $R/\Rsun$
& $L/(10^3\Lsun)$
& $\Teff/(10^3\,{\rm K})$
& $\Tc/(10^7\,{\rm K})$ \\
\midrule
RM1 & 0.0 & 2.00 & 0.080 & 12.2 & 2.40 \\
RM2 & 0.8 & 536  & 20.8  & 2.99 & 8.73 \\
RM3 & 0.9 & 609  & 25.2  & 2.95 & 8.99 \\
RM4 & 1.0 & 679  & 29.6  & 2.91 & 9.21 \\
\bottomrule
\end{tabular}
\caption{Characteristics of the four reference models RM1-RM4 calculated with \mesa{}. All models have $\Mb=3\,\Msun$ and $\RD=0.02\,\Rsun$.}
\label{tab:models}
\end{table}

In Fig.~\ref{fig:mesahr}(b), we also compare the numerical results with those obtained using the homology approximation of Section~\ref{sec:homology}, assuming electron-scattering opacity and nuclear burning through the CNO cycle (Case~D of Table~\ref{tab:homregimes}). Both calculations predict a shift towards lower effective temperatures in the HR diagram. However, they give quite different results for the luminosity at small dark core masses. For instance, at $\MD\simeq0.1\,\Msun$, the homology approximation gives $L\simeq91\,\Lsun$ and $\Teff\simeq5700\,{\rm K}$, while the numerical calculation gives $L\simeq20\,\Lsun$ and $\Teff\simeq5300\,{\rm K}$. In particular, the homology approximation does not reproduce the minimum in the luminosity. The agreement improves at larger dark core masses. For $\MD=0.2$, $0.3$, $0.5$, and $1.0\,\Msun$, the luminosities obtained with the two methods differ by less than $25\%$. However, in the homology calculation we assume a constant effective temperature of $3700\,{\rm K}$ along the cool branch, whereas in the numerical calculation the effective temperature decreases from about $4200\,{\rm K}$ at $\MD=0.2\,\Msun$ to $2900\,{\rm K}$ at $\MD=1\,\Msun$. The homology approximation therefore describes the displacement towards the region of cool and luminous stars in the HR diagram reasonably well, but cannot reproduce all the changes in the stellar envelope.

\begin{figure}[t!]
 \centering
 \includegraphics[width=\linewidth]{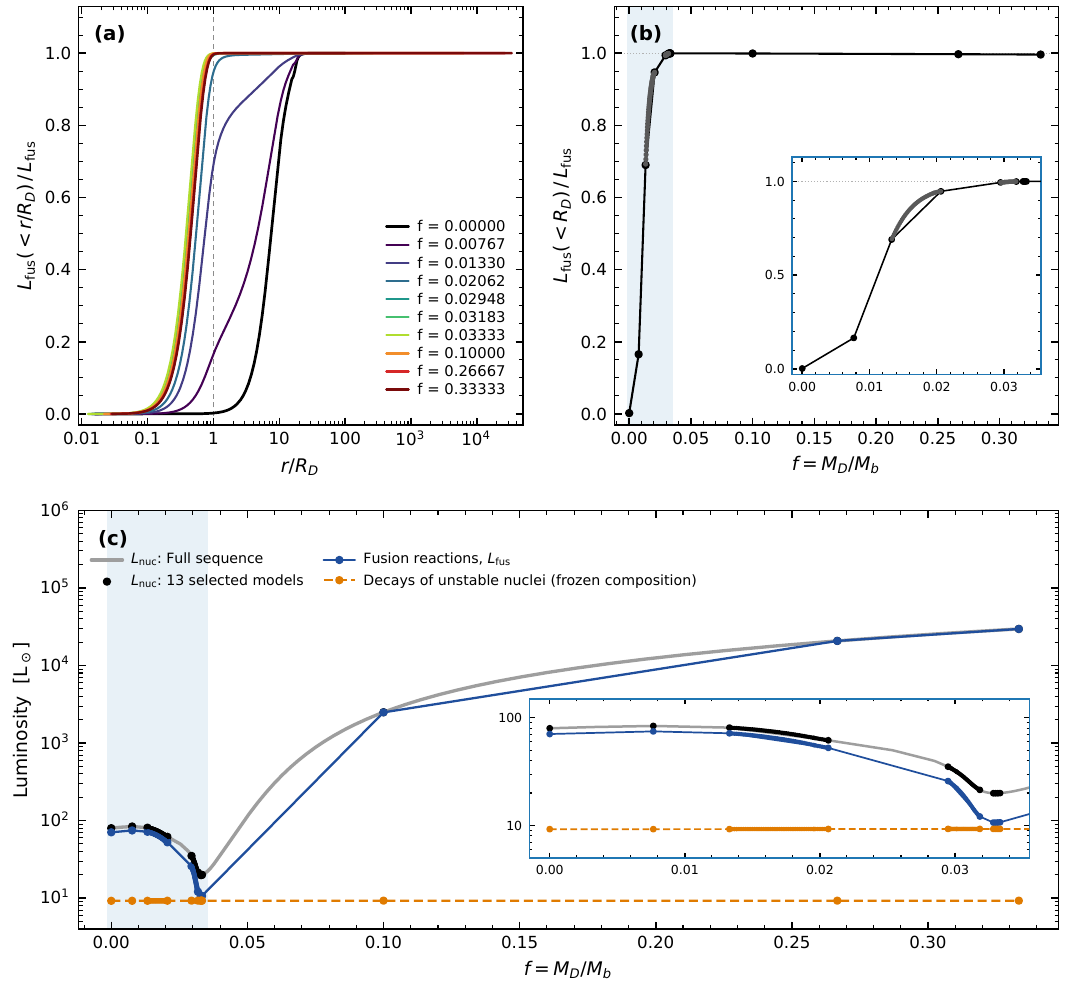}
 \caption{Nuclear energy production for the equilibrium configurations obtained with \mesa{}. Panel (a) shows the cumulative luminosity from nuclear fusion as a function of radius, panel (b) shows the fraction of this luminosity produced within $\RD$, and panel (c) shows the contributions from the different nuclear processes.}
 \label{fig:burning}
\end{figure}

To understand why the luminosity reaches a minimum and why the star later expands, we look more closely at the regions where nuclear burning and convection take place. We write the total nuclear luminosity as $L_{\rm nuc}=L_{\rm fus}+L_{\rm dec}$, as shown in Fig.~\ref{fig:burning}, with
\begin{equation}\label{eq:fusion_budget}
 L_{\rm fus}=\int_0^{\Mb}\epsilon_{\rm fus}(\rho,T,\{X_i\})\,dm_b.
\end{equation}
Here, $L_{\rm fus}$ is the contribution from fusion reactions, while $L_{\rm dec}$ is the contribution from the decay of unstable nuclei included in the reaction network. The quantity $\epsilon_{\rm fus}$ is the corresponding energy-generation rate per unit baryonic mass, and $X_i$ are the nuclear mass fractions. All of this energy is produced by ordinary matter, including the gas inside the dark core.

As the dark core becomes more massive, nuclear burning is pushed into the region inside $\RD$. Figure~\ref{fig:burning}(a) shows the cumulative fusion luminosity, while panel (b) shows the fraction produced at $r<\RD$. In the star without a dark core, only $0.25\%$ of the fusion luminosity is generated inside $\RD$. This fraction increases to about $95\%$ at $f=0.0206$ and becomes larger than $99\%$ for $f\gtrsim0.0295$. The gravitational attraction of the dark core therefore concentrates nuclear burning in a very small central region. However, this does not mean that the total luminosity must increase. In fact, Fig.~\ref{fig:burning}(c) shows that the fusion luminosity can become smaller even when a larger fraction is produced inside $\RD$.

\begin{figure}[t!]
 \centering
 \includegraphics[width=\linewidth]{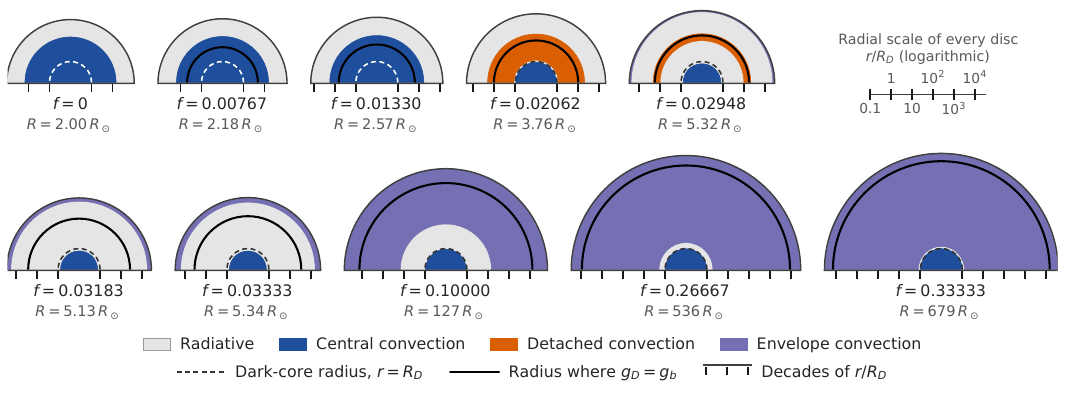}
 \caption{Convective and radiative regions in representative \mesa{} equilibria, shown as logarithmically scaled half-discs. Blue indicates central convection, orange a detached convective zone, purple envelope convection, and gray radiative layers. The black semicircles indicate the radius at which $M_D(r)=m_b(r)$.}
 \label{fig:convection}
\end{figure}

As nuclear burning becomes more concentrated near the center, the convective regions of the star also change significantly, as shown in Fig.~\ref{fig:convection}. Without a dark core, the star has a convective core that extends up to $r\simeq15\,\RD$ and contains about $19\%$ of the baryonic mass. At $f\simeq0.0195$, a radiative layer appears around $r=\RD$, splitting the convective core into two separate convective regions. As $f$ increases further, the outer of the two convective regions becomes smaller and disappears at $f\simeq0.0301$. At the same time, convection starts to develop in the outer layers of the star and gradually moves inward. For $f\geq0.10$, more than $99\%$ of the baryonic mass is inside the convective envelope. At $f=1/3$, the inner edge of this envelope is at about $1.3\,\RD$. As the dark core mass increases, the region where the dark component dominates the gravitational field also becomes larger. The black semicircles in the figure show the radius at which the gravitational pull from dark matter and baryonic matter is the same, that is, where $M_D(r)=m_b(r)$. The result is a star with a very small nuclear-burning region at the center and a large, cool convective envelope around it.

The luminosity reaches its minimum during this transition. Between $f\simeq0.0295$ and $0.0318$, the central temperature decreases from $3.72\times10^7$ to $3.45\times10^7\,{\rm K}$, while the central density increases by about $36\%$. Over the same interval, the luminosity produced by nuclear fusion drops from about $26$ to $12\,\Lsun$, even though the fraction produced within $\RD$ increases slightly from $99.4\%$ to $99.9\%$. Most of the energy comes from the CNO cycle, which depends very strongly on temperature. This explains why the luminosity decreases even though the density increases. The detached convective region also disappears in the same interval of $f$.

When interpreting the minimum in the luminosity, one should keep in mind that the abundances of all nuclei, including unstable ones, are kept fixed in our calculation. As a result, the luminosity from nuclear decays remains approximately constant, $L_{\rm dec}\simeq9.3\,\Lsun$. At $f\simeq0.033$, this accounts for about $47\%$ of the total nuclear luminosity. Allowing the abundances of unstable nuclei to change with temperature and density could therefore affect the results near the luminosity minimum. At larger values of $f$, however, the contribution from nuclear decays becomes negligible.

We now compare the results obtained with \mesa{} with those of the homology approximation presented in Section~\ref{sec:homology}. We focus on Cases B and D, since nuclear burning in the star without a dark core is dominated by the CNO cycle. For each quantity $Q=R,L,\Tc,\Teff$, we define the relative difference
\begin{equation}\label{eq:homology_deviation}
 \delta_Q(f)\equiv
 \frac{Q_{\rm num}(f)/Q_{\rm num}(0)}
 {Q_{\rm hom}(f)/Q_{\rm hom}(0)}-1,
\end{equation}
where $Q_{\rm num}$ and $Q_{\rm hom}$ are the results obtained with \mesa{} and the homology approximation, respectively. In both calculations, the quantities are normalized to their values for the star without a dark core.

\begin{figure}[t!]
 \centering
 \includegraphics[width=0.95\linewidth]{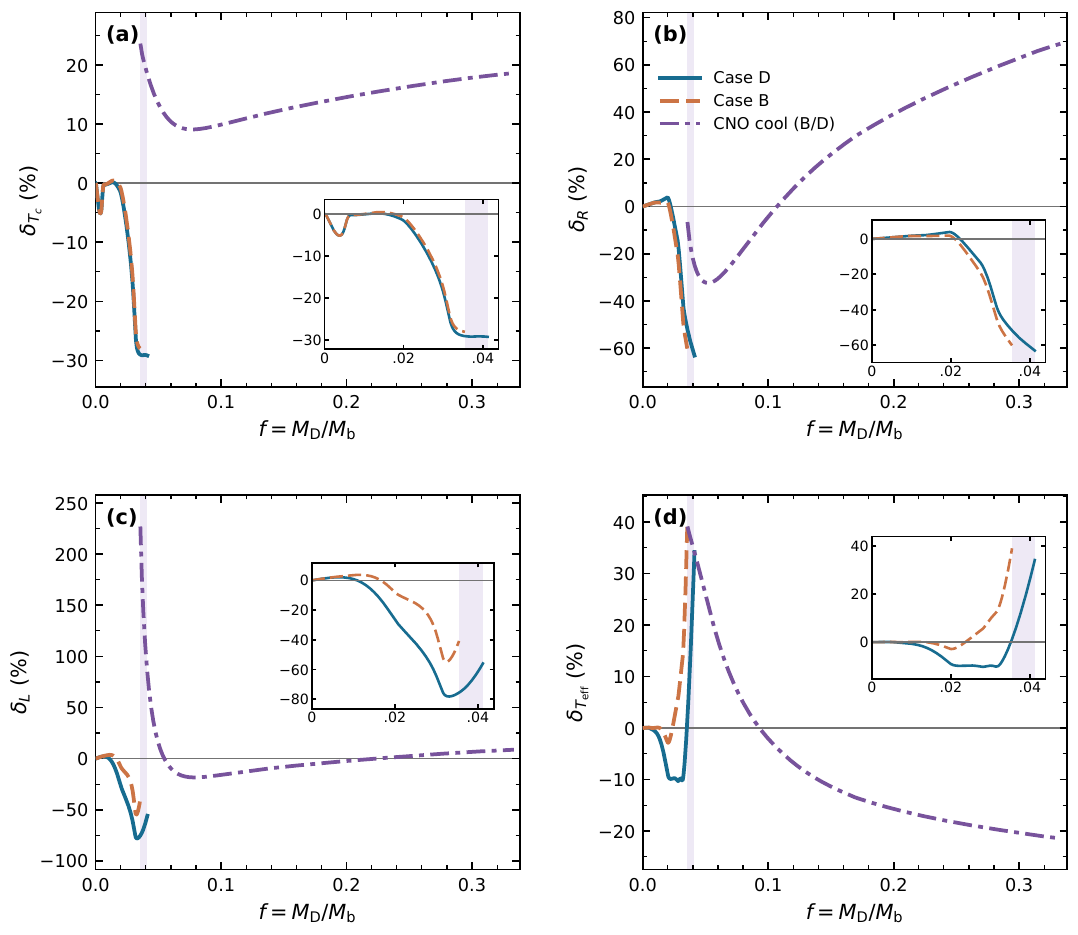}
 \caption{Relative differences between the results obtained with \mesa{} and the homology approximation, as defined in Eq.~\eqref{eq:homology_deviation}, for the central temperature (top left), radius (top right), luminosity (bottom left) and effective temperature (bottom right). The solid, dashed and dash-dotted lines correspond to Cases D, B and their common CNO cool branch, respectively. The insets show the results along the radiative branch in more detail.}
 \label{fig:homology_test}
\end{figure}

As shown in Fig.~\ref{fig:homology_test}, the two calculations agree well for small values of $f$. In Case D, the luminosity is the first quantity for which the relative difference exceeds $10\%$. This occurs at $f\simeq0.0206$, among the values of $f$ considered in the numerical calculation. The homology approximation does not reproduce the minimum in the luminosity, as it does not describe the changes in the convective regions of the star and the associated decrease in central temperature and energy production through the CNO cycle. Instead, Case D predicts a monotonic increase in the luminosity along the radiative branch, given by $L/L_0=(1+f)^4$.

The shaded region in the figure indicates the interval between the transitions from the radiative branch to the cool branch in Cases B and D, which occur at $f\simeq0.0355$ and $0.0413$, respectively. At each transition, both branches give the same effective temperature, $\Teff=T_{\rm H}=3700\,\mathrm{K}$, but different values of the radius, luminosity and central temperature. This explains the discontinuities in the relative differences for $R$, $L$ and $\Tc$. These discontinuities are due to the way we match the two branches in the homology approximation and do not correspond to sudden changes in the structure of the star.

At $f=0.247$, the relative differences between the numerical results and those of the homology approximation along the common CNO cool branch are about $51\%$ for the radius, $2\%$ for the luminosity, $16\%$ for the central temperature and $-18\%$ for the effective temperature. It is interesting that the two calculations give very similar results for the normalized luminosity. However, the effective temperature obtained with \mesa{} is lower than the value $T_{\rm H}$ assumed in the homology approximation, which explains the larger stellar radius.

\begin{figure}[t!]
 \centering
 \includegraphics[width=0.95\linewidth]{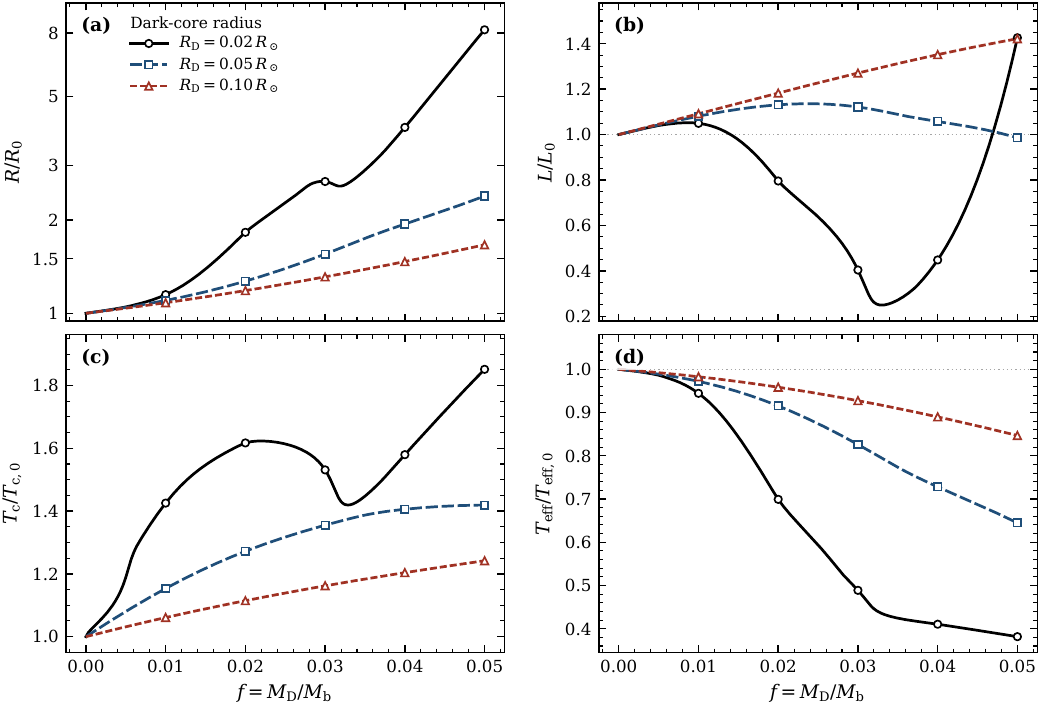}
 \caption{Stellar properties obtained with \mesa{} as functions of $f$ for three different dark core radii $\RD$. The stellar radius, luminosity, central temperature and effective temperature are normalized to their values for the star without a dark core.}
 \label{fig:core_radius}
\end{figure}

The effects of the dark core on the stellar structure also depend on its radius, as expected from Eqs.~\eqref{eq:lambda0} and \eqref{eq:boost}. To study this dependence, we repeat the \mesa{} calculation for three different dark core radii, $\RD=0.02$, $0.05$, and $0.10\,\Rsun$, varying $f$ between zero and $0.05$ and keeping all other physical assumptions unchanged. As shown in Fig.~\ref{fig:core_radius}, for all nonzero values of $f$ considered, a more compact dark core leads to a higher central temperature, a larger stellar radius and a lower effective temperature. At $f=0.05$, the stellar radii obtained for the smallest and largest dark core radii differ by about a factor of five. The behavior of the luminosity is more complicated. For $\RD=0.10\,\Rsun$, the luminosity increases over the whole range of $f$, while for $\RD=0.05\,\Rsun$ it reaches a maximum at $f\simeq0.024$ and then decreases. For $\RD=0.02\,\Rsun$, the luminosity goes through the minimum discussed above and then increases rapidly, becoming larger than that for $\RD=0.05\,\Rsun$ at $f\simeq0.047$. At $f=0.05$, the luminosities obtained for $\RD=0.02$ and $0.10\,\Rsun$ are almost the same. These results show that the stellar properties depend not only on the amount of dark matter inside the star but also on its distribution.

\begin{figure}[t!]
 \centering
 \includegraphics[width=0.95\linewidth]{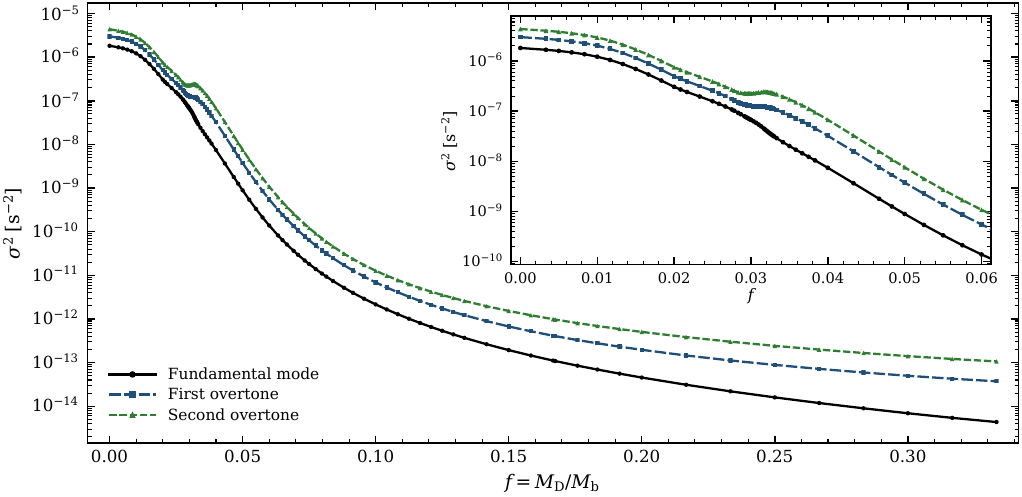}
 \caption{Squared frequencies of the fundamental radial mode and the first two overtones for the equilibrium configurations obtained with \mesa{} for $\RD=0.02\,\Rsun$. All three modes have positive squared frequencies for the configurations considered in the range $0\leq f\leq1/3$. The inset shows the results at small values of $f$ in more detail.}
 \label{fig:radial_stability}
\end{figure}

Finally, we study the stability of the equilibrium configurations obtained with \mesa{} against small adiabatic radial perturbations, assuming that the dark core remains fixed. For $\RD=0.02\,\Rsun$, we calculate the squared frequencies $\sigma^2$ of the fundamental radial mode and the first two overtones for $0\leq f\leq1/3$. As shown in Fig.~\ref{fig:radial_stability}, the squared frequencies of all three modes are positive for all the configurations considered. The stars are therefore stable against linear adiabatic radial perturbations under our assumptions.

The squared frequency of the fundamental mode decreases monotonically from $1.82\times10^{-6}$ to $4.35\times10^{-15}\,{\rm s^{-2}}$ as $f$ increases from zero to $1/3$. This is mainly because the mean density of the star decreases as the star expands. Near $f\simeq0.03$, the squared frequency of the second overtone exhibits a local minimum and a local maximum. Interestingly, these occur in the same region of $f$ where the stellar structure and the convective regions undergo significant changes.

The numerical results show that the effects of a compact dark core cannot be described by simply rescaling the properties of a star without a dark core. The additional gravitational attraction changes the internal structure of the star, concentrating nuclear burning near the center and modifying the regions where energy is transported by convection. For sufficiently large dark core masses, the star develops an extended and cool convective envelope. As a result, a star with an unevolved chemical composition can acquire the appearance of a red giant.
\FloatBarrier
\section{Conclusions}
\label{sec:conclusions}

We have investigated the effects of a compact dark matter core on the structure of a star, assuming that dark matter interacts with ordinary matter only through gravity. We first studied the problem using the homology approximation and then performed a more detailed numerical analysis with \mesa{}. The homology approximation provides a simple description of how the additional gravitational field affects nuclear burning and energy transport, but becomes inaccurate when the stellar structure departs significantly from that of an ordinary star.

For a star with baryonic mass $\Mb=3\,\Msun$ and a dark core of radius $\RD=0.02\,\Rsun$, we find that increasing the dark mass can lead to a dramatic expansion of the stellar envelope. For $\MD=1\,\Msun$, the stellar radius increases from approximately $2$ to $680\,\Rsun$, while the luminosity rises from $80$ to $3\times10^4\,\Lsun$. At the same time, the effective temperature decreases from $1.2\times10^4$ to $2900\,{\rm K}$, despite the increase in central temperature. The star thus moves from the main sequence into the region of the HR diagram occupied by cool, luminous giants, while retaining its original chemical composition.

This behavior results from the concentration of nuclear burning near the dark core and the reorganization of convection. As the dark mass increases, an increasing fraction of the nuclear energy is produced within the dark sphere, while a deep convective envelope develops. The luminosity also exhibits a non-monotonic dependence on the dark mass, which is not reproduced by the homology approximation. 

Our results show that a compact dark object can significantly alter the observable properties of a star through its gravitational effects alone. In particular, a star with an unevolved chemical composition can have the appearance of a red giant. Its position in the HR diagram is therefore not necessarily indicative of its stage of nuclear evolution.

\section*{Acknowledgments}
This work is supported by the Collaborative Research Center SFB1258 and by the Deutsche Forschungsgemeinschaft (DFG, German Research Foundation) under Germany's Excellence Strategy - EXC-2094 - 390783311. H.S. thanks Mattia Di Mauro and INFN Torino for their kind hospitality.

\section*{Note Added}
While finalizing this work, we became aware of a concurrent study~\cite{BetancourtKamenetskaia:2026gqt} discussing a similar idea.

\bibliographystyle{unsrt}
\bibliography{references}
\end{document}